\documentclass[preprint,notitlepage,nofootinbib,superscriptaddress]{revtex4}
\usepackage{ulem}
\usepackage{amssymb, amsmath, amsopn, color, graphicx}
\usepackage[english]{babel}

\numberwithin{equation}{section}
\renewcommand\theequation{\arabic{section}.\arabic{equation}}

\usepackage{bm}
\DeclareMathAlphabet{\matheul}{U}{eus}{m}{n}
\usepackage{float}

\usepackage[colorlinks]{hyperref}
\hypersetup{linkcolor=blue,citecolor=blue,filecolor=black,urlcolor=blue}

\newcommand{\diag}{\mathop{\mathrm {diag}}}

\allowdisplaybreaks

\begin{document}

%\preprint{APS/123-QED}

\title{Physical interpretation of Newman-Janis rotating systems. I. A unique family of Kerr-Schild systems}% Force line breaks with \\
\thanks{}%

\author{Philip Beltracchi}
\email{phipbel@aol.com}
\affiliation{Department of Physics and Astronomy, University of Utah\\Salt Lake City, Utah 84112}

\author{Paolo Gondolo}
\email{paolo.gondolo@utah.edu}
\affiliation{Department of Physics and Astronomy, University of Utah\\Salt Lake City, UT 84112}
\affiliation{Department of Physics, Tokyo Institute of Technology, Tokyo 152-8551, Japan}
\affiliation{Kavli Institute for the Physics and Mathematics of the Universe, The University of Tokyo, Kashiwa, Chiba 277-8583, Japan}

\collaboration{}%\noaffiliation

\date{\today}% It is always \today, today,

\begin{abstract}
\noindent
The Newman-Janis algorithm and its generalizations can be used mathematically to generate rotating solutions from nonrotating spherically-symmetric solutions within general relativity. The energy-momentum tensors of these solutions may or may not represent the same physical system, in the sense of both being a perfect fluid, or an electromagnetic field, or a $\Lambda$-term, and so on. In a series of two papers, we compare the structure of the eigenvalues and eigenvectors of the rotating and nonrotating energy-momentum tensors (their Segre types) and look for the existence of equations of state relating the energy density and the principal pressures. Part I covers Kerr-Schild systems, Part II more general systems. 
 We find that there is a unique family of stationary axisymmetric Kerr-Schild systems that obey the same equation of state in both the rotating and nonrotating configurations. This family includes the Kerr and Kerr-Newman black holes, as well as rotating spacetimes whose mass function in the nonrotating limit contains a constrained superposition of a cloud of strings term, a Reissner-Nordstrom term, a cosmological constant term, and a Schwarzschild term. We describe the common equation of state relating energy density and pressure in this family of spacetimes and discuss some of its properties.
\begin{description}
\item[Usage]

\item[PACS numbers]
 
\item[Structure]
\end{description}
\end{abstract}

\pacs{Valid PACS appear here}% PACS, the Physics and Astronomy
\maketitle

%\tableofcontents
\section{Introduction}

There is a way to find rotating solutions starting from a nonrotating spherically symmetric solution. It is the way Newman, Janis, and collaborators recovered the Kerr solution and discovered the Kerr-Newman solution starting from the Schwarzschild and Reissner-Nordstrom metrics \cite{Newman:1965tw,Newman:1965my,Kerr:2007dk}. The method of Newman and Janis was extended by Gurses and Gursey to all systems of the Kerr-Schild type \cite{Gurses:1975vu}. This Gurses-Gursey generalization has recently been used to derive rotating versions of a variety of systems such as nonsingular black holes \cite{Smailagic:2010nv,Bambi:2013ufa, Ghosh:2014pba,Dymnikova:2015hka, Lamy:2018zvj}, systems with Kiselev ``quintessence" \cite{Ghosh:2015ovj,Sakti:2019iku,Benavides-Gallego:2018odl}, clouds of strings \cite{Sakti:2019iku}, and black holes with nonlinear electrodynamics charge \cite{Atamurotov:2015xfa,Benavides-Gallego:2018odl}, NUT charge \cite{Erbin:2016lzq}, or dilatons \cite{PhysRevD.100.024028}.  The Newman-Janis algorithm was further extended by Drake and Szekeres to create rotating spacetimes from general static spherically symmetric metrics \cite{Drake:1998gf}, a similar scheme with different notation is described in \cite{Benavides-Gallego:2018odl}.

Although the Newman-Janis algorithm creates rotating generalizations of the original metrics,  the physics of the system is in the energy-momentum tensor. We want to find out if the rotating solutions obtained through the Newman-Janis algorithm and its extensions describe the same physical system as in the nonrotating solution but set into rotation. We expect that the Segre type of the nonrotating system is a specialization of the rotating system, for instance, a system that allows the pressures to be different along different axes in the rotating solution may well be in a degenerate state with isotropic pressures in the static spherically symmetric solution. Also, the rotation should cause momentum density terms which can be undone locally with an appropriate comoving boost. Finally, we would expect that any relation between energy, pressure, and stress obeyed by the underlying physical substance can be satisfied in both the rotating and nonrotating stress-energy tensors.

Some methods for modeling rotating solutions within general relativity are designed to specifically preserve these sorts of behaviors.  The Hartle perturbative formalism \cite{Hartle:1967he,Hartle:1968si} is specifically designed to preserve both the equation of state and perfect fluid nature for systems in slow uniform rotation, and more general scenarios of rotating perfect fluids can be treated numerically in the ADM framework \cite{2013rrs..book.....F}. The Newman-Janis algorithm is not specifically designed with the preservation of physical properties in mind, and there are indications that the Newman-Janis system does not always correspond with a physically rotating version of the original spherical system. For instance, the Newman-Janis algorithm does not produce properly rotating monopole fields for Born-Infeld electrodynamics sources \cite{Lombardo_2004}. Additionally, Drake and Szekeres find the only perfect fluid system generated by their method is the vacuum Kerr solution \cite{Drake:1998gf}. In this paper, we systematically examine the energy-momentum tensors of Newman-Janis systems to determine what sort of physical system it represents and when it can be considered a physically rotating version of the original system.
\section{The Newman-Janis method}
The original Newman-Janis algorithm involved writing the Schwarzschild or Reissner-Nordstrom metric in advanced null coordinates, 
\begin{equation}
    ds^2=-\bigg(1-\frac{2M}{r}-\frac{Q^2}{r^2}\bigg)du^2-2\, du\, dr+r^2(d\theta^2+\sin^2 \theta d\phi^2) ,
\end{equation}
expressing $g^{\mu\nu}$ from null tetrads, 
\begin{equation}
    g^{\mu\nu}=-l^\mu n^\nu -l^\nu n^\mu+\bar{m}^\nu m^\mu+\bar{m}^\mu m^\nu,
\end{equation}
and allowing the coordinate $r$ to take complex values such that
\begin{subequations}
\begin{align}
&    l^\mu=\delta^\mu_1 , \\
&    n^\mu=\delta^\mu_0-\frac{1}{2}\left(1-m\Big(\frac{1}{r}+\frac{1}{\bar{r}}\Big)-\frac{Q^2}{r\bar{r}}\right)\delta^{\mu}_1 , \\
&    m^\mu=\frac{1}{\sqrt{2}\bar{r}}\Big(\delta^\mu_2+\frac{i}{\sin \theta} \, \delta^\mu_3\Big).
\end{align}
\end{subequations}

Here an overbar denotes complex conjugation. Then performing a complex transformation on the $r$ and $u$ coordinates to new $r^*$, $u^*$ coordinates,
\begin{subequations}
\begin{align}
    u^*=u-i a \cos \theta,\\
    r^*=r+i a \cos \theta
\end{align}
\end{subequations}
(here the star denotes new coordinates and not complex conjugation), computing the new tetrad $l^{*\mu}$, $n^{*\mu}$, $m^{*\mu}$, $\bar{m}^{*\mu}$, and forming the new metric $g^{*\mu\nu}$, which is real, 
one can take it to be the rotating metric.

One can write an analogous process for arbitrary Kerr-Schild systems. We start with a spherically symmetric Kerr-Schild metric in Schwarzschild coordinates
 \begin{equation}
    ds^2=-\left(1-\frac{2 m(r)}{r}\right)dt^2+\left(1-\frac{2 m(r)}{r}\right)^{-1}dr^2+ r^2 d\theta^2+r^2 \sin^2\theta  d\phi^2.\label{sphmetmet}
\end{equation}
The function $m(r)$, which we will oftentimes simply write as $m$,  may be thought of as giving the enclosed mass in the spherically symmetric system.
Next, we convert to advanced null coordinates $du=dt-(1-2m(r)/r)^{-1}dr$, obtaining
 \begin{equation}
    ds^2=-\bigg(1-\frac{2m(r)}{r}\bigg)du^2-2\, du\, dr+r^2(d\theta^2+\sin^2 \theta d\phi^2).
\end{equation}
This metric can be written in terms of a null tetrad as
\begin{equation}
    g^{\mu\nu}=-l^\mu n^\nu -l^\nu n^\mu+\bar{m}^\nu m^\mu+\bar{m}^\mu m^\nu,
\end{equation}
where in $x^\mu=(u,r,\theta,\phi)$ coordinates we have
\begin{subequations}
    \begin{align}
    l^\mu&=(0,1,0,0),\\
    n^\mu&=\bigg(1,-\frac{1}{2}\Big(1-\frac{2m(r)}{r}\Big),0,0\bigg),\\
    m^\mu&=\bigg(0,0,\frac{1}{\sqrt{2}r},\frac{i}{\sqrt{2}r\sin{\theta}}\bigg).
\end{align}
\end{subequations}
 Notice that $l,n,m,\bar{m}$ are all null, that $l^\mu n_\mu=-1$, and that $m^\mu \bar{m}_\mu=1$.
 Next, we replace $n$ with
 \begin{equation}
     n^\mu=\bigg(1,-\frac{1}{2}\Big(1-\frac{(r+\bar{r})m(\frac{r+\bar{r}}{2})}{r\bar{r}}\Big),0,0\bigg),\\
     \label{newnarbM}
 \end{equation}
 where as earlier $r$ is complex and $\bar{r}$ is its complex conjugate.
 Note that when $r$ is fully real, Eq.~\eqref{newnarbM} reduces to the original $n$. Another important point is that in the original Newman-Janis algorithm, the $M$ term and $Q$ term superficially appear to have been complexified in a different way from each other, but one recovers both terms correctly with the single complexification scheme in Eq.~\eqref{newnarbM}.
 
 After resetting $n$, we change coordinates to $r^*$, $u^*$ with
 \begin{subequations}
 \begin{align}
    u^*=u-i a \cos \theta,\\
    r^*=r+i a \cos \theta,
\end{align}
\end{subequations}
 resulting in a new set of vectors 
\begin{subequations}
 \begin{align}
     l^{*\mu}&=(0,1,0,0), \\
     n^{*\mu}&=\bigg(1,-\frac{1}{2}\Big(1-\frac{2 r^* m(r^*)}{r^{*2}+a^2 \cos^2 \theta}\Big),0,0\bigg) , \\
     m^{*\mu}&=\frac{1}{\sqrt{2}(r^*-i a \cos\theta)}\bigg(i a \sin \theta,-i a \sin \theta,1,\frac{i }{\sin \theta}\bigg) ,
 \end{align}
 \end{subequations}
 if we take $r^*,a,\theta$ to be real.
 Now we construct
 \begin{equation}
    g^{*\mu\nu}=-l^{*\mu} n^{*\nu} -l^{*\nu} n^{*\mu}+\bar{m}^{*\nu} m^{*\mu}+\bar{m}^{*\mu} m^{*\nu},
\end{equation}
and relabel $r^*\rightarrow r$ for simplicity, to obtain
\begin{align}
    ds^2 &=
    -\bigg(1-\frac{2rm}{\Sigma}\bigg) \, du^2
    -2 \, du \, dr
    +\Sigma \, d\theta^2
    \nonumber \\ & 
    +\sin^2\theta\bigg[
    2a \, dr \, d\phi
    - \frac{4 a r m}{\Sigma} \, du \,d\phi
    +\bigg(r^2+a^2+\frac{2 a^2 r m \sin^2 \theta}{\Sigma}\bigg)\, d\phi^2
    \bigg]
\end{align}
with the notation
\begin{align}
\Sigma(r,\theta)=r^2+a^2\cos^2\theta.
\end{align} 
Finally converting to Boyer-Lindquist coordinates using
\begin{subequations}
\begin{align}
    & du=dt-\frac{r^2+a^2}{\Delta}\, dr , \\
    & d\phi=d\varphi-\frac{a}{\Delta}\, dr
\end{align}
\end{subequations}
with the notation
\begin{align}
     \Delta(r)=r^2+a^2-2r m(r),
\end{align}
and relabeling $d\varphi\rightarrow d\phi$, we get
\begin{align}
   \begin{split}
    ds^2=-\left(1-\frac{2 r m}{\Sigma}\right)dt^2+\frac{ \Sigma }{\Delta } \, dr^2+ \Sigma \, d\theta^2+\sin ^2\theta \left(\frac{2 a^2 r  m \sin ^2\theta }{\Sigma }+a^2+r^2\right)d\phi^2\\-\frac{4 a  r  m \sin ^2\theta }{\Sigma } \, dt \, d\phi
    .\label{spinmet}
   \end{split}
\end{align}
which is our rotating metric.

Since we are interested in the properties of the energy momentum tensors rather than the steps of the algorithm itself, we can simply consider the Eq.~(\ref{spinmet}) is the rotating version of the static metric Eq.~(\ref{sphmetmet}). When $a\rightarrow0$, metric (\ref{spinmet}) reproduces the metric (\ref{sphmetmet}), so we may think of $a$ as the rotation parameter. 

 It is helpful to rearrange terms to call attention to the principal directions in Eq.~(\ref{spinmet}),
\begin{align}
ds^2 = - \frac{\Delta}{\Sigma} \big( dt - a \sin^2\theta \,d\phi \big)^2 + \frac{\Sigma}{\Delta} \, dr^2 + \Sigma \, d\theta^2 + \frac{\sin^2\theta}{\Sigma} \big[ (r^2+a^2)d\phi - a \, dt \big]^2 .\label{principal}
\end{align}

Metrics of the Kerr-Schild class can be written $g_{\mu \nu}=\eta_{\mu \nu}-S k_\mu k_\nu$, where $S$ is a scalar function and $k^\mu$ is a null vector with respect to both $g_{\mu \nu}$ and $\eta_{\mu \nu}$ \cite{Stephani:2003tm}.   To put Eq.~(\ref{sphmetmet}) into explicit Kerr-Schild form, introduce the new coordinate $t^*$, satisfying
 \begin{equation}
     dt^*=dt+dr-\left(1-\frac{2 m}{r}\right)^{-1}dr,
 \end{equation}
 and obtain
 \begin{equation}
     ds^2=-(d t^*) ^2+dr^2+r^2 \, d\theta^2+r^2 \sin^2\theta \, d\phi^2+\frac{2 m}{r}(dt^*-dr)^2.
 \end{equation}
 The first four terms are the Minkowski metric, and $k_\mu dx^\mu=dt^*-dr$. Any spherically symmetric Kerr-Schild metric may be put into this form.
 For  Eq.~(\ref{spinmet}), the Kerr-Schild vector and scalar function in Boyer-Lindquist coordinates (see e.g. \cite{Gurses:1975vu})\footnote{Our vector $k^\mu$ is their vector $\lambda^\mu$ in the example after their equation (4.17), and our function $S$ is twice their function $V$ in their Eq. (4.17). } are
 \begin{subequations}
 \begin{align}
    &  S=\frac{2r m}{r^2+a^2 \cos^2\theta} , \\
    & k_\mu=\left(1,\frac{\Sigma}{\Delta},0,-a \sin^2 \theta\right).
     \label{njks}
 \end{align}
 \end{subequations}
 In these coordinates, it may be verified that the remaining portion
 \begin{align}
     \eta_{\mu \nu}dx^\mu dx^\nu=-dt^2+\frac{ \Sigma  \left(a^2-4 r
   m+r^2\right)}{\Delta ^2} \, dr^2+\Sigma \, d\theta^2+
   \left(a^2+r^2\right) \sin ^2\theta \, d\phi^2+\nonumber\\\frac{4 a
    r m \sin ^2\theta}{\Delta }\, dr \, d\phi -\frac{4 r m}{\Delta } \, dr \, dt
 \end{align}
 has a fully 0 Riemann tensor, so it is the required flat space portion of the Kerr-Schild metric.  
  
 We will refer to physical systems having rotating Kerr-Schild metrics of the form in Eqs.~\eqref{spinmet} or~\eqref{principal} as Gurses-Gursey rotating systems, with a corresponding nonrotating metric in Eq.~\eqref{sphmetmet}.

\section{Nonrotating Kerr-Schild systems}

The energy-momentum tensor for a static spherical Kerr-Schild system \eqref{sphmetmet} has nonzero mixed components
\begin{align}
T^t_{~t} = T^r_{~r} = - \frac{m'}{4\pi r^2} , \qquad T^\theta_{~\theta} = T^\phi_{~\phi} = - \frac{m''}{8\pi r} ,
\label{sphereThl}
\end{align}
where a prime denotes a derivative with respect to $r$.
 The eigenvalues of the nonrotating energy-momentum tensor~(\ref{sphereThl}) can easily be obtained since Eq.~(\ref{sphereThl}) is already diagonal,
    \begin{align}
     -\rho^0 = p_\parallel^0 =-\frac{ m'}{4 \pi r^2}, \qquad
     p_\perp^0= -\frac{m''}{8 \pi r} .
     \label{unisphereeigs}
 \end{align} 
Here $\rho^0$ is the energy density, $p_\parallel^0$ is the principal pressure in the radial direction $r$,  and $p_\perp^0$ is the principal pressure in the tangential directions $\theta$ and $\phi$.

The Segre type of nonrotating energy-momentum tensor~(\ref{sphereThl}) is [(11)(1,1)].  The possible degenerate case [(111,1)] occurs when $-\rho^0 = p_\perp^0$, which is equivalent to
 \begin{equation}
     m''=\frac{2m'}{r} .
 \end{equation}
The solution to the latter equation is 
 \begin{equation}
     m(r)= \frac{\Lambda}{6} r^3+M,
     \label{sdsmass}
 \end{equation}
 which in general describes the Schwarzschild/de Sitter spacetime. When $\Lambda=0$, the energy-momentum tensor is $T^\mu_{~\nu}=0$, and its Segre type is properly 0, but for simplicity we will refer to it as being of [(111,1)] type. 

In the rest of the section, we examine spherically symmetric Kerr-Schild systems in the context of their equations of state.
 One useful quantity for equations of state for the spherical systems is the isotropic
 pressure $p$ defined by
 \begin{equation}
    p=\frac{p_r+2p_T}{3},    \label{piso}
\end{equation}
 using the common notation $p_r=p_\parallel^0$ and $p_T=p_\perp^0$ for the radial and tangential pressure of static spherically symmetric systems. It is important to note, because this has caused confusion \cite{Visser:2019brz} about the Kiselev solution \cite{Kiselev:2002dx}, that the existence of this quantity and usage of this quantity in an equation of state {\it does not} imply that the medium has isotropic pressure.
 
 Spherically symmetric Kerr-Schild systems automatically satisfy one equation of state $p_r=-\rho$. If a system is defined with a second equation of state $p_T(\rho)$ or $p(\rho)$ rather than the mass function, then deriving the mass function involves solving one of the differential equation systems
\begin{subequations}
\label{mvspT}
 \begin{align}
  \frac{d^2m}{dr^2}= -8\pi r \, p_T(\rho),  \qquad
  \rho=\frac{1}{4\pi r^2}\frac{dm}{dr},
\intertext{or}
\frac{d^2m}{dr^2}+\frac{1}{r}\frac{dm}{dr}=-12\pi r \,  p(\rho), \qquad
\rho = \frac{1}{4\pi r^2}\frac{dm}{dr}.
\end{align}
\end{subequations}

There is ample choice of mass functions $m(r)$ in the literature, or equivalently of equations of state $p_T(\rho)$ or $p(\rho)$. As we will see in Sec.~\ref{sec:eos}, most of these choices do not preserve the equation of state in passing from the nonrotating to the rotating system by means of the Newman-Janis algorithm. Nevertheless, we have collected some notable mass functions and corresponding equations of state for static spherically-symmetric Kerr-Schild spacetimes in the Appendix, in particular the class of linear equations of state $p=w\rho$ with constant $w$, and the nonsingular black hole solutions of Bardeen~\cite{Bardeen}, Hayward~\cite{Hayward:2005gi}, and Dymnikova~\cite{Dymnikova1992}.

\section{Rotating energy-momentum tensors}

For a rotating Gursey-Gurses system \eqref{spinmet},  the nonzero components of the energy-momentum tensor are
\begin{subequations}
\label{bigspinsT}
  \begin{align}
 & T^t_{~t}= - \frac{m'}{4\pi\Sigma^3} \left[ r^2(r^2+a^2)-a^4\sin^2\theta\cos^2\theta\right] + \frac{ra^2\sin^2\theta \, m''}{8\pi\Sigma^2},\\[1ex]
& T^\phi_{~t}= \frac{a}{8\pi\Sigma^3} \left[ (r^2+a^2\cos^2\theta) r m'' -2 (r^2-a^2\cos^2\theta) m' \right] ,\\[1ex]
& T^r_{~r}= -\frac{r^2 m'}{4 \pi  \Sigma ^2} , \\[1ex]
& T^\theta_{~\theta}= - \frac{rm''}{8\pi\Sigma} - \frac{a^2 \cos ^2\theta \, m'}{4 \pi  \Sigma ^2} , \\[1ex]
& T^t_{~\phi}= - \sin ^2\theta \, (a^2+r^2) \,\, T^\phi_{~~t}  , \\[1ex]
& T^\phi_{~\phi}= \frac{a^2m'}{4\pi\Sigma^3} \left[ r^2\sin^2\theta-(r^2+a^2)\cos^2\theta\right]  - \frac{r(a^2+r^2) m''}{8\pi\Sigma^2}  .
\end{align}
 \end{subequations}
 The eigenvalues  $\Lambda_{(i)}$ ($i=1,\ldots,4$) of the rotating energy-momentum tensor~(\ref{bigspinsT}) are obtained by diagonalization. One finds two pairs of degenerate eigenvalues, $\Lambda_{(1)}=\Lambda_{(2)}$ and $\Lambda_{(3)}=\Lambda_{(4)}$, with the eigenspace of one pair having a (normalized) timelike eigenvector $\tilde{u}^\mu$ and giving the comoving energy density $\rho=T_{\mu\nu} \tilde{u}^\mu \tilde{u}^\nu$ and principal parallel pressure $p_\parallel$ as 
 \begin{subequations}
 \label{spinDENP}
  \begin{align}
     &-\rho = p_\parallel = \Lambda_{(1)} = \Lambda_{(2)}=-\frac{r^2 m'}{4\pi  \Sigma^2} ,
     \label{spinDEN}
\intertext{and the other pair giving the principal perpendicular pressure $p_\perp$ as}
     & p_\perp = \Lambda_{(3)}=\Lambda_{(4)}=- \frac{rm''}{8\pi \Sigma} - \frac{a^2 \cos ^2\theta \, m'}{4 \pi  \Sigma ^2}.
     \label{spinP}
 \end{align}
 \end{subequations}
Details on the eigenvectors and covariant decomposition of this energy-momentum tensor are explored in Sec.~\ref{sec:invariant_spaces}.
 
 In terms of these eigenvalues, the covariant energy conservation equation $\nabla_\mu T^\mu_{~\nu}=0$ in the rotating Gurses-Gursey spacetime~\eqref{spinmet} becomes
\begin{subequations}
\label{eq:energy_conservation}
   \begin{align}
 &   \frac{\partial \rho}{\partial r}= - \frac{2r}{\Sigma} \, (p_\perp+\rho), \\
&    \frac{\partial p_\perp}{\partial \theta}=\frac{a^2 \sin 2\theta}{\Sigma} \, (p_\perp+\rho) .
\end{align}
\end{subequations}
 In the nonrotating $a\rightarrow0$ limit, using the common notation $p_r=p_\parallel^0$ and $p_T=p_\perp^0$ for the radial and tangential pressure of static spherically symmetric systems, these become $ {\partial \rho}/{\partial r}={-2(p_T+\rho)}/{r}$, which is the anisotropic Tollman-Oppenheimer-Volkov equation~\cite{1974ApJ...188..657B} with $p_r=-\rho$, and ${\partial p_T}/{\partial \theta}=0$, which is a consequence of the spherical symmetry of the system.

\subsection{Segre types}

Regarding the Segre type of the energy-momentum tensor, since the eigenvalues come in two pairs of degenerate eigenvalues, Gurses-Gurses rotating systems are of Segre type [(11)(1,1)] or its degenerate case [(111,1)], just like the corresponding nonrotating systems. Further, we
 see that the superposition behavior present in spherical systems \cite{Gurses:1975vu,Ibohal:2004kk, 2019arXiv191008166B} is also preserved in the sense that the eigenvalues and all the components of the energy-momentum tensor (\ref{bigspinsT}) are linear equations in $m$ such that combination systems may be obtained simply by adding $m$ functions. 
 The degenerate case $p_\perp=-\rho$ requires $m(r)=M$, and no longer includes the de Sitter cosmological term in $\Lambda$ of the degenerate nonrotating case in Eq.~\eqref{sdsmass}.
 
As covariant eigenvectors $v^{(i)}_\mu$ of Eq.~(\ref{bigspinsT}) in the $t,r,\theta,\phi$ coordinates satisfying
\begin{align}
v^{(i)}_\mu T^\mu_{~\nu} = \Lambda_{(i)} \, v^{(i)}_\nu \qquad (i=1,\ldots,4),
\end{align}
we can take
\begin{subequations}
\label{eq:v_covectors}
     \begin{align}
    & v^{(1)}_\mu=(-1,0,0,a \sin^2 \theta)\\
    & v^{(2)}_\mu=(0,1,0,0)\\
  & v^{(3)}_\mu=(0,0,1,0)\\
   &  v^{(4)}_\mu=\left(-\frac{a}{a^2+r^2},0,0,1\right) .
 \end{align}
  \end{subequations}
Note that these eigenvectors are orthogonal to each other and are the principal directions from Eq.~(\ref{principal}). Due to the degeneracy of the eigenvalues, any linear combination of $v^{(1)}_\mu$ and $v^{(2)}_\mu$, and any linear combination of $v^{(3)}_\mu$ and $v^{(4)}_\mu$, is also an eigenvector. We call the eigenspace of $p_\parallel=-\rho$ spanned by $v^{(1)}_\mu$ and $v^{(2)}_\mu$ at any given spacetime point the ``parallel principal plane'' of the Gursey-Gurses system at that point, and the eigenspace of $p_\perp$  spanned by  $v^{(3)}_\mu$ and $v^{(4)}_\mu$ the ``transverse principal plane.'' 
 
 It is also interesting to examine how the rotating eigenvalues $-\rho=p_\parallel$ and $p_\perp$ are related to the nonrotating eigenvalues $-\rho^0=p^0_\parallel$ and  $p^0_\perp$ in Eq.~(\ref{unisphereeigs}). The rotating eigenvalues are a linear function of the nonrotating eigenvalues,
 \begin{align}
    -\rho=-\frac{r^4 \rho^0}{\Sigma^2} , \qquad
    p_\perp=\frac{r^4 \rho^0}{\Sigma^2}+\frac{r^2 (p_\perp^0-\rho^0)}{\Sigma} .
     \label{spineigswitheigs}
 \end{align}
This suggests the possibility that the rotating system is some kind of linear deformation of the nonrotating system. 

For the special case of Schwarzschild/de Sitter spacetimes in Eq.~\eqref{sdsmass}, which have degenerate Segre type [(111,1)] with nonrotating eigenvalues $p^0_\parallel=p^0_\perp$, the rotating eigenvalues are no longer degenerate if the cosmological term $\Lambda\ne 0$,
  \begin{align}
  p_\perp - p_\parallel = - \frac{\Lambda r^2 a^2 \cos^2\theta}{4\pi\Sigma^2} ,
 \end{align}
as follows from Eqs.~\eqref{spinDENP} for $m(r)$ in Eq.~\eqref{sdsmass}.
 These ``rotating de Sitter'' spacetimes are examined in Sec.~\ref{sec:rotating_deSitter}. The degenerate case [(111,1)] with $\Lambda=0$ is the Kerr spacetime.

 To summarize, the Segre type of a rotating Kerr-Schild system obtained via the Newman-Janis algorithm in the way of Gurses and Gursey is [(11)(1,1)], the same as the Segre type of the nonrotating system. However, the degenerate [(111,1)] Segre type in the rotating case includes only the Kerr spacetime, which originates from the Schwarzschild term of the nonrotating degenerate [(111,1)] spacetimes. The other nonrotating Kerr-Schild systems of degenerate Segre type [(111,1)], namely those with a nonzero cosmological term in \eqref{sdsmass}, become Gurses-Gursey rotating systems with a nondegenerate Segre type [(11)(1,1)].

 In the rest of this section, we further describe the eigenvector/eigenvalue structure of the energy-momentum tensor  $T^{\mu}_{~\nu}$ for Gurses-Gursey rotating systems.

 \subsection{\boldmath Invariant spaces of $T^{\mu}_{~\nu}$ and angular velocity}
\label{sec:invariant_spaces}
 
  A covariant version of Eq.~(\ref{bigspinsT}) is the spectral decomposition, common to all spacetimes of Segre type [(11)(1,1)], 
 \begin{align}
 T_{\mu\nu}=p_\parallel L_{\mu \nu}+p_\perp H_{\mu \nu} \label{twithvecs},
 \end{align}
where the tensor $L^{\mu}_{~\nu}$ is a projector onto the eigenspace of $p_\parallel=-\rho$, which we called the parallel principal plane, and the tensor $H^{\mu}_{~\nu}=\delta^{\mu}_{~\nu}-L^{\mu}_{~\nu}$ is the projector onto the eigenspace of $p_\perp$, which we called the transverse principal plane. For these projectors, $L^{\mu}_{~\lambda} L^{\lambda}_{~\nu} = L^{\mu}_{~\nu}$, $H^{\mu}_{~\lambda} H^{\lambda}_{~\nu} = H^{\mu}_{~\nu}$, $L^{\mu}_{~\lambda} H^{\lambda}_{~\nu}=0$. The parallel principal plane contains both spacelike and timelike vectors and has the structure of a two-dimensional Minkowski spacetime. In particular, it contains two independent null directions, which are the principal null directions of the energy-momentum tensor (or equivalently of the Ricci tensor). On the other hand, the transverse principal plane is composed of spacelike vectors only.  

The projector onto the parallel principal plane may be written in terms of two independent null vectors $k_\mu$ and $l_\mu$ along the two independent lightlike directions in it as
 \begin{align}
     L_{\mu\nu}=\frac{k_\mu l_\nu+k_\nu l_\mu}{k^\alpha l_\alpha}.
 \end{align}
The lightlike directions, i.e., the principal null directions of the Ricci tensor, are uniquely defined within the parallel tangent plane, although the normalization of the vectors $k_\mu$ and $l_\mu$ is arbitrary, all normalizations giving the same projector $L^{\mu}_{~\nu}$. For fixed $k^\alpha l_\alpha$, the remaining choice of normalization amounts to a local Lorentz transformation in the parallel principal plane, which leaves the principal null directions invariant. 
For the Gurses-Gurses system in Eq.~\eqref{spinmet},  $k_\mu$ can be taken to be the Kerr-Schild vector defined in Eq.~\eqref{njks} and $l_\mu$ as the independent principal null vector $l_\mu=(1,-\frac{\Sigma}{\Delta},0,-a \sin^2 \theta)$. 

The projector onto the parallel principal plane can also be written in terms of a timelike vector and a spacelike vector belonging to it, which can, in particular, be chosen to be orthonormal.  Previous authors (e.g.~\cite{Dymnikova:2015yma}) have used
 \begin{align}
&     L_{\mu\nu}=\mathop{\rm sign}(\Delta)\, (-u_\mu u_\nu +d_\mu d_\nu)
\end{align}
with
\begin{align}
&     u^\mu=\frac{1}{\sqrt{|\Delta| \Sigma }}\big[ (r^2+a^2)\delta^\mu_t+a\delta^\mu_\phi\big],\qquad d^\mu=\sqrt{\frac{| \Delta|}{\Sigma}}\, \delta^\mu_r,\label{uvec}
 \end{align}
 which they liken to the covariant form for an ``anisotropic fluid." There is a similar decomposition presented in the original Gurses-Gursey paper \cite{Gurses:1975vu}, but with a missing square root in the normalization. 
 Note that the $u^\mu$ and $d^\mu$ vectors defined here are scaled versions of our eigenvectors $v^{(1)\mu}$ and $v^{(2)\nu}$. In this form, $u^\mu$ is suggestive of a four velocity (in a region where $t$ is timelike and $\Delta>0$) with nonzero $t$ and $\phi$ components. However, this four velocity is not unique. One may equivalently use vectors
 \begin{align}
 \label{eq:tilde_u}
     \tilde{u}^\mu=\cosh(W) u^\mu+\sinh(W) d^\mu,\qquad\tilde{d}^\mu=\cosh(W) d^\mu+\sinh(W) u^\mu,
 \end{align} 
 to construct $L^{\mu\nu}$ because of the invariance of Segre type [(11)(1,1)] systems under Lorentz boosts in the parallel principal plane spanned by $u^\mu$ and $d^\mu$. 
 
 The same invariance can be understood using local Lorentz frames. We can express the energy-momentum tensor in a local Lorentz frame using the tetrad
 \begin{align}
     e^\mu_{~\hat{\mu}}=\left(
\begin{array}{cccc}
 \frac{a^2+r^2}{\sqrt{|\Delta|  \Sigma }} & 0 & 0 &
   \frac{a \sin \theta}{\sqrt{\Sigma}} \\
 0 & \sqrt{\frac{|\Delta|
 }{\Sigma }} & 0 & 0 \\
 0 & 0 & \sqrt{\frac{1}{\Sigma }} & 0 \\
 \frac{a}{\sqrt{|\Delta|  \Sigma }} & 0 & 0 &
   \frac{1}{\sin \theta \sqrt{\Sigma} } \\
\end{array}
\right), 
\label{tet1}
 \end{align}
where the spacetime index $\mu$ labels the rows and the orthonormal index $\hat{\mu}$ labels the columns. The vectors $e^\mu_{~\hat{\mu}}$ are the normalized version of the contravariant vectors $v^{(i)\mu}$ corresponding to Eq.~\eqref{eq:v_covectors}. Note that if $\Delta$ is negative, then the $r$ coordinate is timelike and the appropriate orthonormal metric is $g_{\hat{\mu}\hat{\nu}}=\diag(1,-1,1,1)$ rather than the usual $g_{\hat{\mu}\hat{\nu}}=\diag(-1,1,1,1)$ which applies when $\Delta>0$.
 With the tetrad~\eqref{tet1}, we find that for $\Delta>0$,
 \begin{align}
     T_{\hat{\mu}\hat{\nu}}=T_{\mu\nu}e^\mu_{~\hat{\mu}}e^\nu_{~\hat{\nu}}=\diag(\rho,-\rho,p_\perp,p_\perp),\label{TLL}
 \end{align}
while for $\Delta<0$, $T_{\hat{\mu}\hat{\nu}}=\diag(-\rho,\rho,p_\perp,p_\perp)$.
 Therefore Eq.~\eqref{tet1} is associated with a special local Lorentz frame in which the energy-momentum tensor is diagonal. Because this is a local Lorentz frame, we may examine the energy-momentum tensor in another local Lorentz frame in motion with respect to the frame in \eqref{tet1} by taking standard Lorentz boosts. We see that boosts in the $\hat{0}\hat{1}$ plane, which is the parallel principal plane, do not change $T_{\hat{\mu}\hat{\nu}}$ in Eq.~(\ref{TLL}), nor do rotations in the $\hat{2}\hat{3}$ plane, which is the transverse principal plane. These symmetries are characteristic of systems with Segre type [(11)(1,1)]. A consequence of this symmetry is the ambiguity in defining a unique four-velocity $u^\mu$ for such systems that we have already seen in Eq.~\eqref{eq:tilde_u}.
 
 Regardless of this ambiguity, one can define a coordinate angular velocity $d\phi/dt$ by eliminating the proper time $d\tau$ from $\tilde{u}^\mu = dx^\mu/d\tau$,
 \begin{equation}
  \frac{d\phi}{d t}=\frac{\tilde{u}^\phi}{\tilde{u}^t}= \frac{a}{a^2+r^2}.
  \label{4vq}
 \end{equation}
This $d\phi/dt$ should not be confused with the frame-dragging angular velocity $\omega=-g_{t\phi}/g_{\phi\phi}$.  The coordinate angular velocity describes a differential rotation, with $d\phi/dt$ which goes as $a/r^2$ for $r\gg a$ and goes to $1/a$ for $r\ll a$. Note that the dependence of  $d\phi/dt$ on $r$ does not depend on $m(r)$ at all, so all Gurses-Gursey rotating systems have the same coordinate angular velocity for a given $a,r$ regardless of their physical content.

\section{Equations of State}
\label{sec:eos}

 In general, equations of state may be written as a function containing thermodynamic variables such as density, pressures, temperature, etc., as
 \begin{equation}
     F(\rho,p,T,...)=0.
     \label{speos_i}
 \end{equation}
 These thermodynamic variables may depend on position and time. However, 
we do not consider functions of the sort 
 \begin{align}
     f(\rho,p,...,x^\mu)=0
     \label{preeos}
 \end{align}
 to be equations of state due to the explicit dependence on position.
 %, but if the dependence on position can be made implicit through relation with a thermodynamic variable then the function can be re-expressed as an equation of state.
 
All Kerr-Schild systems we consider are Segre type [(11)(1,1)] and automatically satisfy the simple equations of state $\rho=-p_\parallel$, $p_2=p_3=p_\perp$.  We are interested in when the system can also satisfy an equation of state of the form
 \begin{equation}
     F(\rho,p_\perp)=0.
     \label{speos}
 \end{equation}
 We are especially interested in when systems have the same equation of state $F(\rho,p_\perp)$ whether they are rotating or not. 
 
We should not expect that all systems should satisfy an equation of state as simple as Eq.~\eqref{speos}. Certain systems may satisfy a more complicated equation of state, for example involving temperature $F(\rho,p_\perp,T)$. Another factor that may be relevant for certain situations is that Kerr-Schild systems may be superposed (by adding their $m$ functions). It is possible to have multiple component systems in which each component satisfies an equation of state of the form~\eqref{speos} but the combined system does not. In such a case, it should be possible to derive a more complicated equation of state for the combined system with additional thermodynamic variables related to the fraction of the total system at a spacetime point which may be ascribed to each individual component.
 
For nonrotating spherical Kerr-Schild systems $p_\perp=p_\perp(r)$ and $\rho=\rho(r)$, so a function of the form in Eq.~\eqref{preeos}, $f(\rho,p_\perp,r)=0$, may be defined. If either $p_\perp(r)$ or $\rho(r)$ is invertible, such that $r=r(\rho)$ or $r=r(p_\perp)$, then the position dependence in $f$ can be eliminated and an equation of state of the form $\rho=\rho(p_\perp)$ or $p_\perp=p_\perp(\rho)$ may be derived. If either $\rho$ or $p_\perp$ is monotonic over a domain $r_{\rm min} \le r \le r_{\rm max}$, then an equation of state defined in this manner applies within that domain. If $r_{\rm min}\rightarrow0$ and $r_{\rm max} \rightarrow \infty$, then the equation of state applies everywhere. If $\rho={\rm const}$ over some domain, then the expressions in terms of $m$ from Eq.~(\ref{sphereThl}) dictate that $p_\perp$ is minus the same constant, and the equation of state $p_\perp=-\rho$ applies. The degenerate Segre [(111,1)] cases, being Minkowski, Schwarzschild, de Sitter, and Schwarzschild/de Sitter, all satisfy the equation of state $p_\perp=-\rho$ globally. In the Appendix, we examine Kerr-Schild systems which follow a linear equation of state and present the equation of state for some nonsingular black hole models.
 
For rotating Gursey-Gurses systems, while the Segre type [(11)(1,1)] is preserved in the standard Newman-Janis algorithm, the relationship between the eigenvalues is, in general, not.  
  One can see this especially from Eq.~(\ref{spineigswitheigs}). The relationship between $\rho^0$ and $\rho$ is straightforward, but $p_\perp$ depends on $p_\perp^0$ and $\rho^0$, with a different functional dependence on $\Sigma$ (or $\theta$) between the terms in general. However, when $p^0_\perp=\rho^0$, the $1/\Sigma$ term in $p_\perp$ goes to 0 and both have the same $\theta$ dependence. Further, we see that if $p_\perp^0=\rho^0$, then $p_\perp=\rho$.  If we use Eq.~(\ref{mvspT}) with the nonrotating equation of state $p_\perp^0=\rho^0$, we obtain $rm''(r)+2m'(r)=0$, or $m=M-k/r$. There are four examples which have this mass function and hence preserve their equation of state: Minkowski to Minkowski $(k=M=0)$, Schwarzschild to Kerr $(k=0, M\ne0)$, Reissner-Nordstrom to Kerr-Newman $(k\ne0,M\ne0)$, and the massless charged particle case $(k\ne0,M=0)$. The Kerr and Kerr-Newman are conventionally viewed as the ``physically correct" rotating versions of black holes. Further, they are the cases for which the equation of state is obviously unmodified by the Newman-Janis algorithm. 
 
 \subsection{A special family of stationary axisymmetric spacetimes}
 \label{sec:family}

  We find that there is a larger class of solutions for which an equation of state exists for Gurses-Gursey rotating systems.  If we take derivatives of $F(\rho,p_\perp)$ in Eq.~\eqref{speos} with respect to $r$ and $\theta$, we obtain
 \begin{subequations}
   \label{Eoscheck}
    \begin{align}
     \frac{\partial F}{\partial\rho} \frac{\partial \rho}{\partial r}+\frac{\partial F}{\partial p_\perp} \frac{\partial p_\perp}{\partial r}=0,\\
      \frac{\partial F}{\partial\rho} \frac{\partial \rho}{\partial \theta}+\frac{\partial F}{\partial p_\perp} \frac{\partial p_\perp}{\partial \theta}=0.
 \end{align}
 \end{subequations}
If this condition is not satisfied, then the Gurses-Gursey system does not satisfy an equation of state of the form $F(\rho,p_\perp)=0$. In order for Eq.~(\ref{Eoscheck}) to hold  for all $r,\theta$ in nontrivial situations, we require that
 \begin{equation}
 \label{eq:eos_condition}
     \frac{\partial p_\perp}{\partial r}\frac{\partial \rho}{\partial \theta}-\frac{\partial p_\perp}{\partial \theta}\frac{\partial \rho}{\partial r}=0
 \end{equation}
 for all $r$ and $\theta$.

 Using the expressions for the eigenvalues Eqs.~(\ref{spinDENP}) in Eq.~\eqref{eq:eos_condition}, we obtain a differential equation for $m$
 \begin{equation}
     r^2 m''(r)^2-2 r m'(r) \left(m''(r)+r
   m^{(3)}(r)\right)+4 m'(r)^2 =0,
   \label{NJeos1}
 \end{equation}
 having the general solution
\begin{subequations}
\label{eq:family}
\begin{align}
m(r)=M-\frac{Q^2}{r}+\lambda r+\frac{1}{6} \Lambda r^3
\label{spinmeos}
\end{align}
with the constraint
\begin{equation}
    \lambda^2=2Q^2 \Lambda.
    \label{eosconstraint}
\end{equation}
\end{subequations}
Here $Q^2$ may be negative when $m(r)$ is considered as a solution of \eqref{NJeos1}, and thus both positive and negative values of $\Lambda$ are acceptable. If $\Lambda=0$, then the constraint~\eqref{eosconstraint} forces $\lambda=0$ and one recovers the Kerr-Newman solution for $Q^2>0$. As we can see from Eq.~(\ref{eosconstraint}), it is not possible to have a system with only $\lambda\ne0$ satisfy an equation of state in the rotating case, there must a $\Lambda$ and $Q$ present.

The static density corresponding to the solution~\eqref{eq:family} is
\begin{align}
\rho^0 = \frac{Q^2}{4\pi r^4} + \frac{\lambda}{4\pi r^2} + \frac{\Lambda}{8\pi}  
= \frac{(\lambda+r^2 \Lambda)^2}{8\pi\Lambda r^4} ,
%= \frac{\Lambda}{8\pi} \left(1+\frac{\lambda}{\Lambda r^2}\right)^2, 
\end{align}
where the second equality uses the constraint~\eqref{eosconstraint} with $\Lambda\ne0$.
The rotating system has energy density $\rho$ and pressures $p_\parallel$ and $p_\perp$ given for $\Lambda\ne 0$ by
\begin{subequations}
\label{eq:rho_family}
\begin{align}
& \rho =  - p_\parallel = \frac{(\lambda+r^2 \Lambda)^2}{8\pi\Lambda (r^2+a^2\cos^2\theta)^2}  ,   \\
& p_\perp = \frac{(\lambda+r^2 \Lambda) ( \lambda-r^2\Lambda-2a^2 \Lambda \cos^2\theta)}{8\pi \Lambda  (r^2+a^2\cos^2\theta)^2}. 
\end{align}
\end{subequations}

The null energy condition $\rho+p_\perp\ge 0$ imposes an interesting maximum value for the rotation parameter $a$. In fact, Eqs.~\eqref{eq:rho_family} give
\begin{align}
\rho + p_\perp = \frac{\lambda+r^2\Lambda}{4\pi\Lambda\Sigma^2} \, ( \lambda - a^2 \Lambda \cos^2\theta ),
\end{align}
which is positive at all $r,\theta$ only if 
\begin{align}
\lambda > a^2 \Lambda > 0 .
\end{align}
Therefore at given values of the family parameters $M,Q,\lambda,\Lambda,a$, the null energy condition is satisfied only if 
\begin{align}
\label{eq:amax}
a \le a_{\rm max} \equiv \sqrt{\lambda/\Lambda} = ( 2Q^2/\Lambda)^{1/4} .
\end{align}

\begin{figure}[t]
    \centering
    \includegraphics[width=10cm]{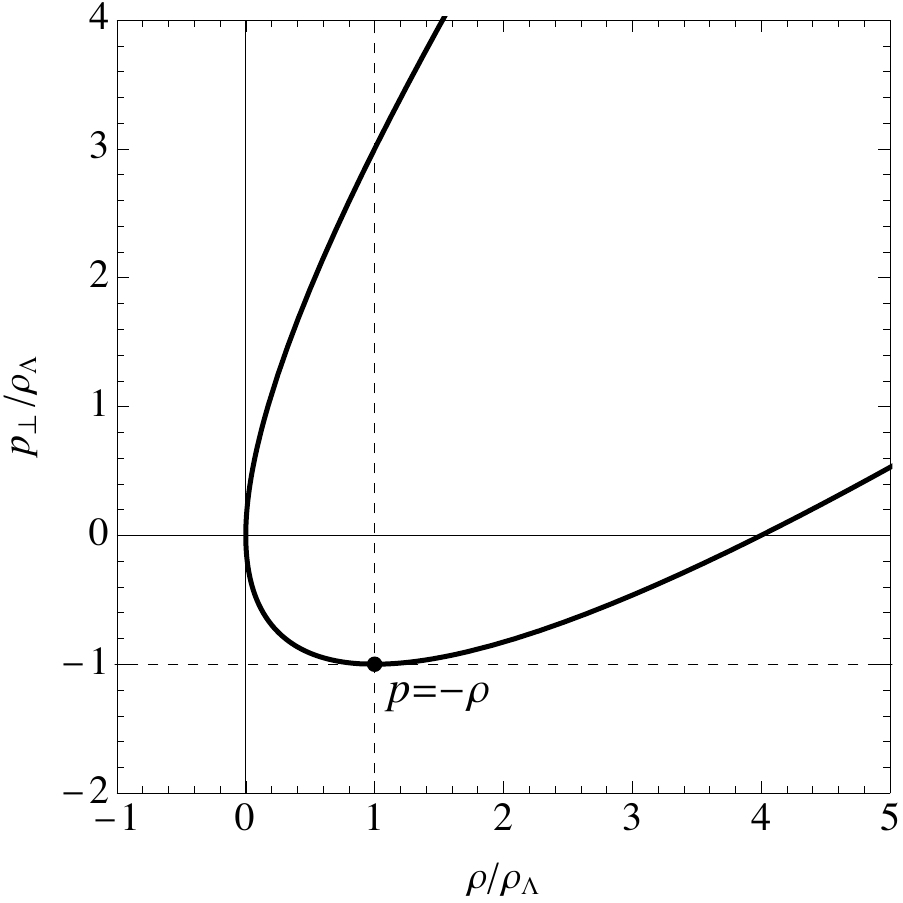}
    \caption{Equation of state $p_\perp(\rho)$ in~\eqref{GGEOS} for the unique family \eqref{eq:family} of rotating Gursey-Gurses systems that preserves the equation of state between rotating and nonrotating systems. The other equation of state is $p_\parallel=-\rho$. The figure assumes  $\Lambda\ne 0$. When $\rho=\rho_\Lambda$ the system satisfies $p_\parallel=p_\perp=-\rho$. At densities $\rho \gg \rho_\Lambda$, the behavior approximates $p_\perp=\rho$,  which is associated with the Reissner-Nordstrom term $Q^2/r$ in $m(r)$.}
    \label{eosplot}
\end{figure}

The equation of state which is satisfied by a system with the $m$ from Eq.~(\ref{eq:family})  in both the rotating and nonrotating case is
\begin{equation}
 (\rho-p_\perp)^2=4 \rho \rho_\Lambda ,
    \label{GGEOS}
\end{equation}
where $\rho_\Lambda=\Lambda/8\pi$. Therefore, a Gurses-Gursey system with a mass function satisfying Eqs.~(\ref{spinmeos}) and (\ref{eosconstraint}) may be interpreted as a physically rotating object made of a substance satisfying Eq.~\eqref{GGEOS}.

Figure \ref{eosplot} shows the equation of state relating $p_\perp$ and $\rho$. The other equation of state is $p_\parallel = - \rho$. Portions of the equation of state surface may be unstable against perturbations, and portions may be stable. An analysis of stability in the case of anisotropic pressures is complicated and is outside the scope of this work.  The equation of state~\eqref{GGEOS} may be of interest for example in cosmology where it may allow for dark energy to be reached dynamically at late cosmic times.

\begin{figure}[t]
    \centering
    \includegraphics[width=10cm]{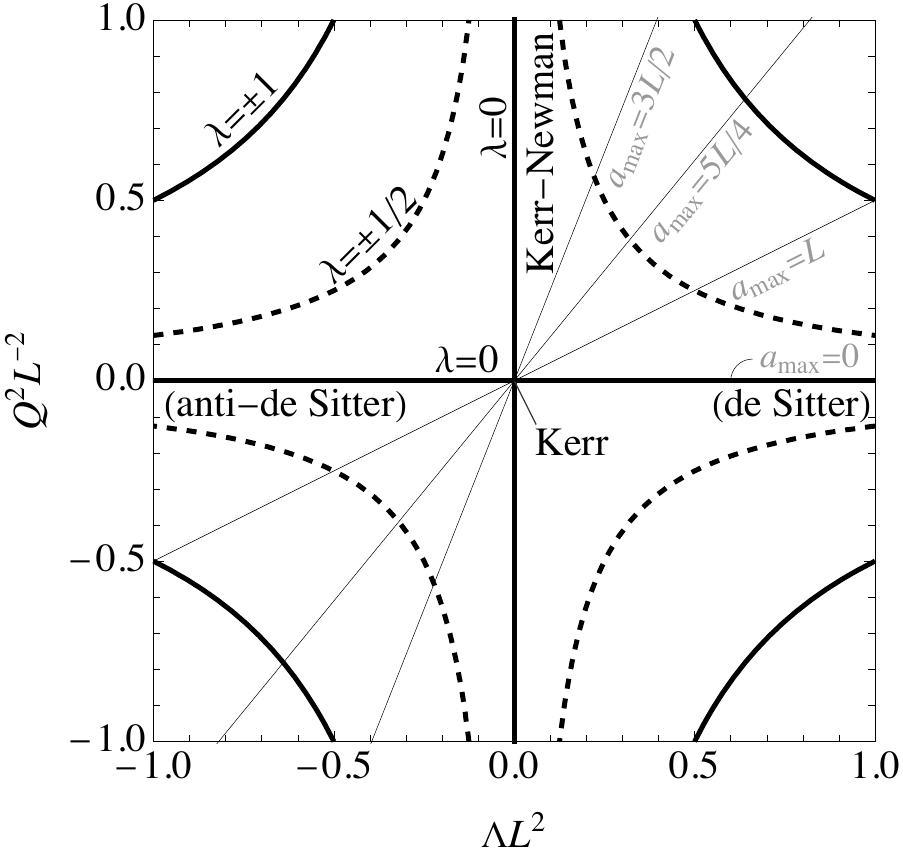}
    \caption{Family of rotating Kerr-Schild spacetimes in the parameters $Q^2,\lambda,\Lambda$ in Eq.~(\protect\ref{eq:family}) in geometrized units, with $L$ an arbitrary unit of length. The Kerr-Newman ($\Lambda=\lambda=0$, $Q^2>0$) and Kerr ($\Lambda=\lambda=Q=0$) limits are marked. The static spherically symmetric limits at $M=\lambda=Q=0$ (de Sitter and anti-de Sitter spacetimes) are indicated in parenthesis. Also indicated are the maximum values $a_{\rm max}$ of the rotation parameter $a$ in Eq.~\protect\eqref{eq:amax} for which the null energy condition $\rho+p_\perp\ge0$ is satisfied.}
    \label{ggEOSsols}
\end{figure}

Note that in the static case, the mass function Eq.~(\ref{spinmeos}) may be interpreted as having a Schwarzschild term with mass $M$, an electromagnetic term with charge $Q$, a global monopole \cite{PhysRevLett.63.341} or string cloud \cite{Letelier:1979ej} term with linear mass density $\lambda $, and a de Sitter term with cosmological constant $\Lambda$.
In the rotating case, the physical interpretation of the system is more subtle. From the Kerr-Newman solution we know that if $\lambda=\Lambda=0$, the rotating system may be thought of as a rotating charged black hole.  If the mass function only has nonzero $\Lambda$, the rotating system does not correspond to what one would expect for rotating vacuum energy for which the equation of state is $\rho=-p_\parallel=-p_\perp$, as the system satisfies the equation of state \eqref{GGEOS} with $\rho \ne - p_\perp$ generally (it is the ``rotating de Sitter'' spacetime discussed in the next Subsection).

Figure~\ref{ggEOSsols} illustrates the parameter space of the family of rotating Kerr-Schild spacetimes in Eq.~\eqref{eq:family}. The parameters $Q^2,\lambda,\Lambda$ are there given in geometrized units in terms of an arbitrary, but common, unit of length $L$. The Kerr-Newman, Kerr, de Sitter, and anti-de Sitter limits are indicated, as well as the maximum values $a_{\rm max}$ of the rotation parameter $a$ for which the null energy condition $\rho+p_\perp\ge0$ is satisfied. Notice in particular that the (anti) de Sitter case satisfies the null energy condition only if they are nonrotating, as they have $a_{\rm max} = 0$. We discuss this case next.

 \subsection{The case of ``rotating de Sitter'' spacetime}
 \label{sec:rotating_deSitter}
 
One illustrative case for rotating Gurses-Gursey solutions obeying the equation of state \eqref{GGEOS} is $M=Q=\lambda=0$, $\Lambda\ne0$. This system is the easiest case that explicitly shows a solution and its nontrivial behavior for systems with the equation of state \eqref{GGEOS}. It shows in particular how the energy density and pressure depend on $r$ and $\theta$ while obeying the equation of state.
This $\Lambda\ne0$ spacetime which results from using the Gurses-Gursey method on de Sitter space has been referred to as ``rotating de Sitter" in Refs.~\cite{Ibohal:2004kk,2006PhLB..639..368D, deUrreta:2015nla} and as a ``rotating imperfect $\Lambda$-fluid" in \cite{Azreg-Ainou:2014nra}. Because spherical Kerr-Schild spacetimes that have a nondivergent Kretschmann scalar at the origin have a de Sitter like core, see Eq.~(\ref{kretchuni2}), ``rotating de Sitter" is found in the cores of the Gurses-Gursey generalizations of these objects \cite{2006PhLB..639..368D,Azreg-Ainou:2014nra}, and has also been examined as limiting case for a model of a Kerr-Newman black hole in a de Sitter background \cite{Ibohal:2004kk,deUrreta:2015nla}. While the ``rotating de Sitter" metric has been considered in various capacities before, the interpretation as being filled with a substance obeying \eqref{GGEOS} seems to be new.

The ``rotating de Sitter" and Kerr systems differ only by replacement of $M$ with $\Lambda r^3/6$ \cite{Azreg-Ainou:2014nra,deUrreta:2015nla}, which is fundamentally because they are both   specific cases of systems of the form in Eq.~(\ref{spinmet}). Using our formulas, we find for the ``rotating de Sitter'' spacetime
 \begin{align}
    \rho=\frac{\Lambda  r^4}{8 \pi \Sigma^2}, \qquad
    p_\perp=-\frac{\Lambda  r^2 (2\Sigma-r^2)}{8 \pi  \Sigma^2}.
 \end{align}
 ``Rotating de Sitter" space has properties which are not de Sitter like. For instance, there is a point $r=0,\theta=\pi/2$ at which the Ricci scalar curvature 
 \begin{equation}
     R=\frac{4 \Lambda  r^2}{r^2+a^2\cos^2\theta}
 \end{equation}
 is undefined, ranging between 0 and $4\Lambda$ depending on the path of approach.
 
 Additionally, the ``rotating de Sitter'' system is Segre type [(11)(1,1)], whereas standard de Sitter is [(111,1)], so the Newman-Janis algorithm has destroyed one degeneracy. Moreover, the ``rotating de Sitter" space violates the null energy condition $ \rho+p_\perp\ge 0 $ everywhere except on the equatorial plane $\theta=\pi/2$, whereas standard de Sitter space satisfies the null energy condition everywhere. Thus the underlying origin of the energy-momentum tensor of the ``rotating de Sitter'' space is not a cosmological constant.
 
If we demand that the vacuum equation of state $p_\perp=p_\parallel=-\rho$ is maintained, and we were to use a vacuum energy-momentum tensor decomposed as in Eq.~(\ref{twithvecs}) in terms of a timelike four-velocity vector as in Eq.~(\ref{uvec}), we find that the vectors $u^\mu$ and $d^\mu$ are arbitrary and don't enter the energy-momentum tensor, so directly spinning  (giving a four-velocity to) vacuum energy does nothing to its energy-momentum tensor. A spinning de Sitter space with vacuum equation of state is described by Carter's solution \cite{Carter2009RepublicationOB} with $M=0$. This solution cannot be reached by the Newman-Janis algorithm, as the only way to obtain $p_\perp=-\rho$ from Eqs.~(\ref{spinDENP}), other than using $a=0$, is $m=M$, and this gives the Kerr spacetime and not Carter's.
 
 However, we can interpret the ``rotating de Sitter" spacetime to be fundamentally filled with a substance that satisfies the equation of state (\ref{GGEOS}) and just happens to be in the special case $\rho=\rho_\Lambda=-p_\perp$ when it is not rotating.

 \section{Conclusions}

 The Newman-Janis algorithm can be used to create the Kerr and Kerr-Newman metrics from the Schwarzschild and Reissner-Nordstrom metrics. Additionally, its generalizations allow for the construction of rotating systems which reduce to spherical systems in the limit of no rotation. 
 
 In the Gurses-Gursey generalization, the rotating systems maintain some properties such as the Segre type [(11)(1,1)], the Kerr-Schild metric class, and the ability to create superimposed systems by adding $m(r)$ functions. Another feature of the Gurses-Gursey rotating systems is that the coordinate angular velocity Eq.~(\ref{4vq}) is fixed in terms of $r,a$ without any dependence on the specific $m(r)$ function in question.
However, for general functions $m(r)$, the relationship between the eigenvalues of the energy-momentum tensor (equations of state) are not preserved in going from the nonrotating to the rotating system. 
 
 We find a unique family of Kerr-Schild systems that maintain the same equation of state in the nonrotating static spherically symmetric case and in the rotating case obtained by means of the Newman-Janis algorithm in the implementation of Gurses and Gursey. This family is described by the mass function $m(r)$ in Eqs.~(\ref{spinmeos}) with the parameters constrained by Eq.~(\ref{eosconstraint}). This family includes the Kerr and Kerr-Newman black holes, obtained through the Newman-Janis algorithm from their corresponding nonrotating Schwarzschild and Reissner-Nordstrom spacetimes, respectively. The other members of the family are rotating spacetimes that correspond in the nonrotating limit to a constrained superposition of the mass functions $m(r)$ of a cloud of strings, the Reissner-Nordstrom spacetime, the (anti) de Sitter spacetime, and the Schwarzschild spacetime. The common equation of state~\eqref{GGEOS} for systems in this family may include stable and unstable configurations, and a more detailed analysis of stability is left for future study.

\acknowledgments

P.G. was partially supported by NSF grant No. PHY-2014075, and is very grateful to Prof.\ Masahide Yamaguchi for his generous support under JSPS Grant-in-Aid for Scientific Research Number JP18K18764 at the Tokyo Institute of Technology.

\bibliographystyle{apsrev4-2}
\bibliography{monster.bib}

\appendix

\renewcommand\theequation{\Alph{section}.\arabic{equation}}

\section{Some equations of state for spherical Kerr-Schild systems}

 In this appendix, we examine some interesting equations of state for static spherically symmetric Kerr-Schild systems. We use the usual notation $p_r=p_\parallel^0$ and $p_T=p_\perp^0$ for the radial and tangential pressure of a static spherically symmetric system which is not necessarily isotropic.
 
\subsection{Linear equation of state}

One simple application of Eq.~(\ref{mvspT}) is showing the connection between linear equations of state and simple power laws for $m$. Within some region, if we have an equation of state of the form
\begin{align}
    p_r=-\rho,\qquad p_T=w_T \rho, \qquad p=w\rho
    \label{eoslin}
\end{align}
where by Eq.~\eqref{piso} we have $w_T=(3w+1)/2$, then the mass will typically be of the form
\begin{align}
    m=\frac{c r^{-3w}}{-3w}+M.\label{powermvw}
\end{align}
This formula had been presented for the Kiselev solution in~\cite{Kiselev:2002dx}, but it applies to any case where the mass follows a power law. 
In the special case $w=0$, the solution to Eq.~(\ref{mvspT})
 becomes
\begin{align}
    m=c \ln(r)+M.
\end{align}
 
 There are several notable example of systems which follow linear equations of state of the form (\ref{eoslin}). 
Minkowski space is a trivial example having $m(r)=0$ and $p=p_T=p_r=-\rho=0$, and de Sitter space has $m(r)=\Lambda r^3/6$, $p=p_T=p_r=-\rho=\Lambda/(8\pi)$. Minkowski and de Sitter are special among these simple cases. For instance, $w_T=w=-1$ applies so the pressure is isotropic everywhere. Additionally, this is the smallest $w_T$ can be and still satisfy the null energy condition
\begin{align}
    \rho+p_T\ge0
\end{align}
and the largest $w_T$ can be such that it remains nonsingular at $r=0$ if $M=0$, as can be seen from the Kretschmann scalar
\begin{align}
    \mathcal{K}=\frac{48 m^2}{r^6}-\frac{64 m m'}{r^5}-\frac{16
   m' m''}{r^3}+\frac{4 m''^2}{r^2}+\frac{4
   \left(8 m'^2+4 m m''\right)}{r^4}
   \label{kretchuni}
\end{align}
which becomes
\begin{align}
   \mathcal{K} =\frac{4 c^2(4+20 w+51 w^2+54 w^3+27 w^4)}{3 w^2 r^{6(1+w)}}
   \label{kretchuni2}
\end{align}
for mass functions of the form (\ref{powermvw}) with $M=0$. A local behavior of $w_T=w=-1$ near $r=0$ allows for regularity and the null energy condition so it is ubiquitous in more complex spherical Kerr-Schild models like nonsingular black holes.
 
 Systems with Segre type [(11)(1,1)] following $p=w\rho$ with $-1<w<-1/3$ are sometimes referred to as quintessence \cite{Kiselev:2002dx}, although this nomenclature is incorrect~\cite{Visser:2019brz}. Kiselev quintessence systems have an infinite total mass $m\propto r^{-3w}$, are not asymptotically flat, and have a de Sitter like outer horizon where $1-2m/r$ changes sign. These systems satisfy the null energy condition but do not satisfy the strong energy condition.

A system $m=\lambda r$, $p_T=0$, $p=-\rho/3$ shows up in different contexts as a collection of radially aligned strings \cite{Letelier:1979ej} or a variety of monopole \cite{PhysRevLett.63.341}. This is a limiting case for the strong energy condition because $\rho+3p=0$. This system is not asymptotically Minkowski (the geometry is hyperconical), and has diverging $m$, but has no de Sitter-like horizon.

 The case  $w=0$ has $m\propto \ln(r)$.  Interestingly, the mass $m$ as $r\rightarrow \infty$ diverges, but the metric is still asymptotically Minkowski because $m/r\rightarrow0$. The origin is singular as can be seen from Eq.~(\ref{kretchuni}).
 
The Schwarzschild black hole has $m(r)=M$, and the pure vacuum equation of state $p=p_T=p_r=\rho=0$. 

One final simple case is the Reissner-Nordstrom solution. If for instance one uses the equation of state for electromagnetism $p=\rho/3$, then $p_T=\rho$, $p_r=-\rho$, and Eq.~(\ref{mvspT}) gives $m=M-Q^2/r$ which is the mass function for the Reissner-Nordstrom solution. The mass function at large radii converges to the constant $M$, but the mass function as $r\rightarrow 0$ diverges. The density follows $\rho\propto r^{-4}$. This is a limiting case for the dominant energy condition, in that $\rho-|p_T|=0$.

Of these spherical Kerr-Schild systems with linear equations of state, the only ones for which the equation of state is preserved under Gurses-Gursey rotation are the electromagnetic $p_T=\rho$, $p=\rho/3$ and its subset the pure vacuum $p=p_T=p_r=\rho=0$.

\subsection{Nonsingular black hole spacetimes}

Here we examine some more complicated spherical Kerr-Schild spacetimes which are used to construct nonsingular black holes.  We derive equations of state for the static  Bardeen, Hayward, and Dymnikova nonsingular black hole spacetimes.

None of these nonsingular black holes have a mass function of the form Eq.~(\ref{eq:family}), so their Gurses-Gursey rotating versions do not satisfy an equation of state $F(\rho,p_\perp)=0$, despite their spherical versions satisfying one. It is still possible that a more general fundamental equation of state involving thermodynamic variables beyond pressure and energy density applies in the rotating case, and that it reduces to the nonrotating equations of state we derive here in the nonrotating case. Finding and justifying more general equations of state for Gurses-Gursey rotating nonsingular black holes is a possible area for future research. For example, since the Hayward and Dymnikova nonsingular black holes can arise from quantum gravity considerations, it would be an interesting avenue to see if quantum gravity considerations allow for the derivation of more general equations of state and whether these equations of state are satisfied by the Gurses-Gursey rotating versions. The Bardeen spacetime has been interpreted as a nonlinear electrodynamics monopole \cite{AyonBeato:2000zs}, but it has been shown that the behavior of nonlinear electrodynamics is not preserved by the Newman-Janis algorithm \cite{Lombardo_2004}, so a different physical explanation for the Bardeen spacetime may be required to allow for derivation of equations of state which apply in both the rotating and nonrotating cases.

The systems in this Section have de Sitter like cores to be nonsingular. Nonsingularity also gives $M=0$, so the system is uniquely defined by an equation of state and has the coordinate $t$ correspond to time for an observer at the origin. Also, all these examples of nonsingular black holes have finite total mass. It is possible for the following spacetimes to lack event horizons for certain parameter ranges, such a system is referred to as a G-lump in  \cite{Dymnikova:2001fb}.

 \subsubsection{Bardeen solution}
 The first nonsingular black hole spacetime discovered was the Bardeen solution \cite{Bardeen}. It was originally proposed as a response to the Penrose singularity theorem \cite{PhysRevLett.14.57}, and has
 \begin{align}
&   m=\frac{M r^3}{(r^2+R^2)^{3/2}},\\
&   \rho=\frac{3 M R^2}{4 \pi (r^2+R^2)^{5/2}},\\
&   p_T=\frac{1}{2} \rho  \left(3-5\left(\frac{\rho}{\rho_0
   }\right)^{2/5}\right), \\
&   p=\frac{1}{3} \rho  \left(2-5\left(\frac{\rho}{\rho_0
   }\right)^{2/5}\right),
 \end{align}
 where $M$ and $R$ are constants and $\rho_0=3M/(4\pi R^3)$. Note that as $\rho\rightarrow\rho_0$ we approach a de Sitter like configuration and as $\rho \ll \rho_0$  the dominant energy condition $\rho-|p_T|\ge0$ is violated, which differs from the implication in \cite{Balart:2014jia}. A plot of the equations of state for the Bardeen spacetime is in Fig. \ref{bardeeneos}.
 
 \subsubsection{Hayward spacetime}
 One popular more recent model of a nonsingular black hole is the Hayward spacetime \cite{Hayward:2005gi}. One reason it has gathered attention because it arises in  ``asymptotically safe quantum gravity" formulations \cite{Saueressig:2015xua}. For the Hayward black hole,
 \begin{align}
&   m=\frac{2M r^2}{r^3+2l^2 M},\\
&   \rho=\frac{3 l^2 M^2}{2 \pi(r^3+2l^2M)^2},\\
&   p_T=\rho\left(2-3\left(\frac{\rho}{\rho_0}\right)^{1/2}\right), \\
&   p=\rho\left(1-2\left(\frac{\rho}{\rho_0}\right)^{1/2}\right),
 \end{align}
where $l$ and $M$ are parameters and $\rho_0=3/(8\pi l^2)$.  A plot of the equations of state for the Hayward spacetime is in Fig. \ref{bardeeneos}.

 \subsubsection{Dymnikova solutions}
 
\begin{figure}[t]
     \centering
     \includegraphics[width=8cm]{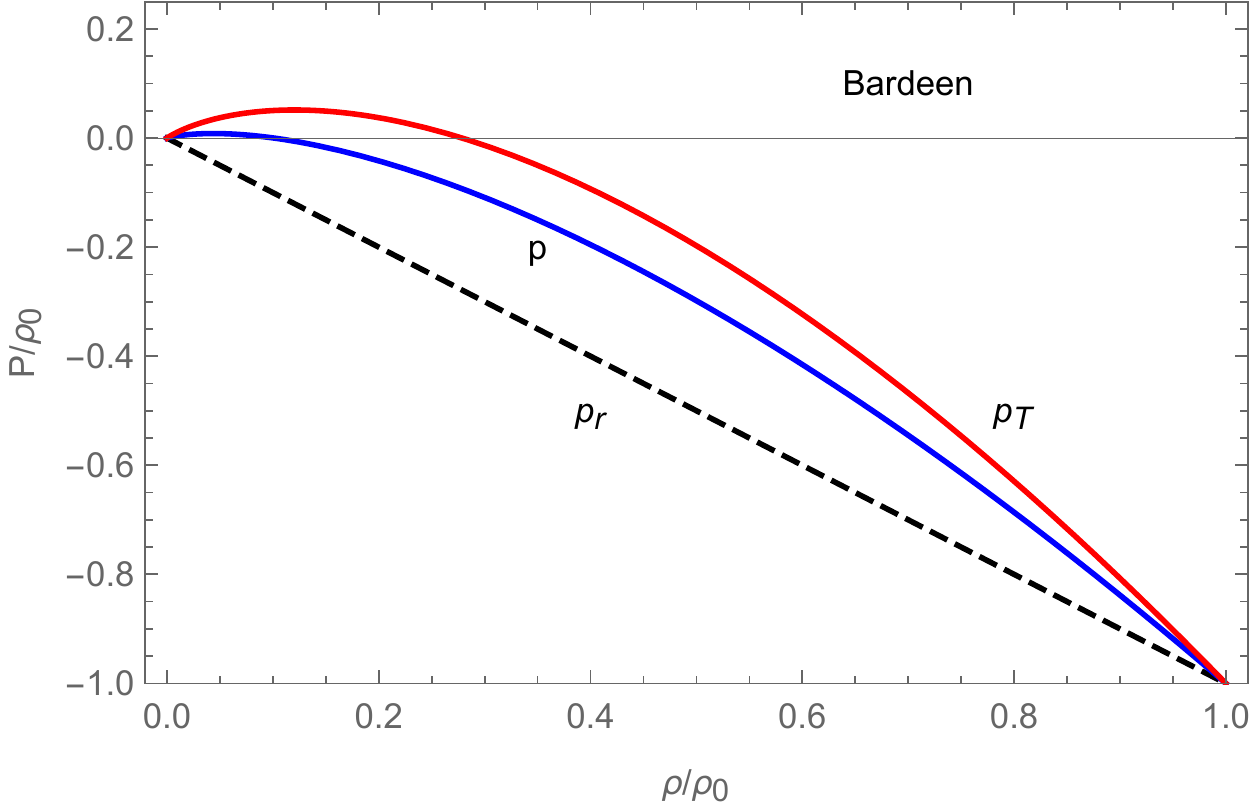}
     \includegraphics[width=8cm]{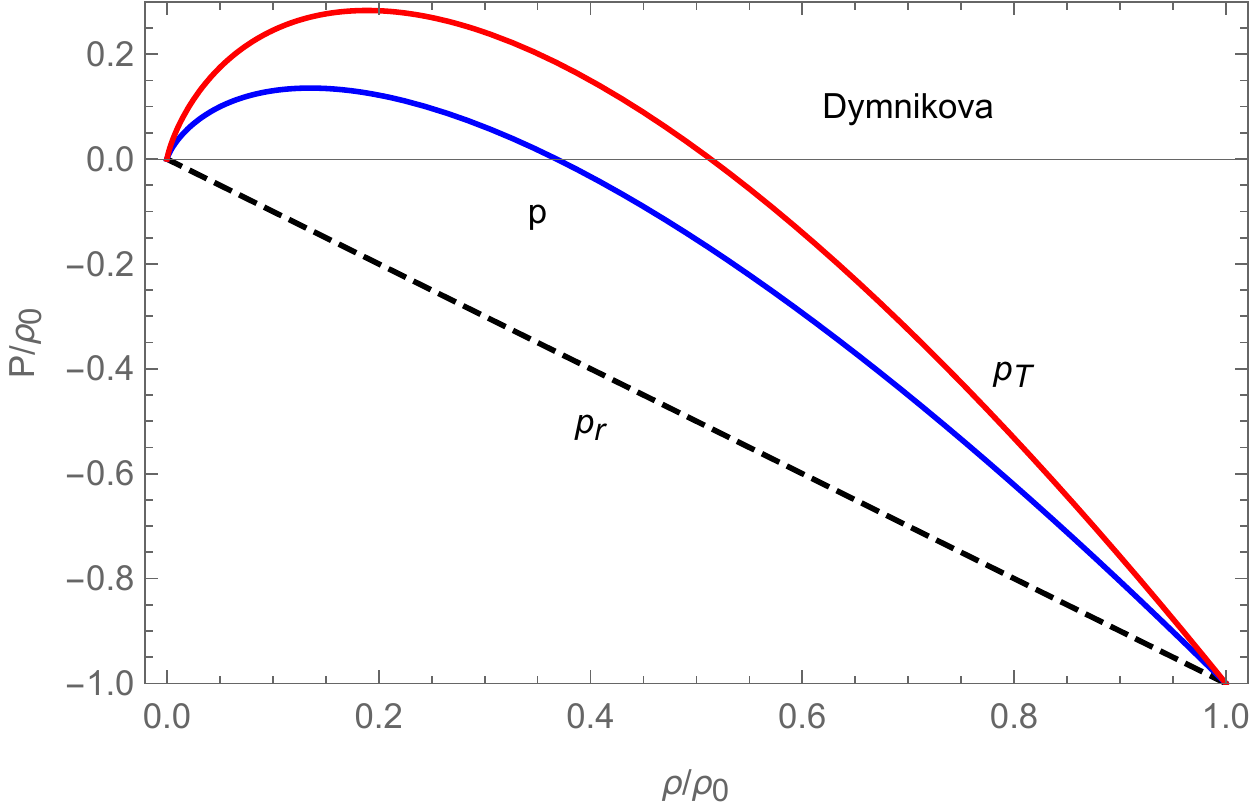}
     \includegraphics[width=8cm]{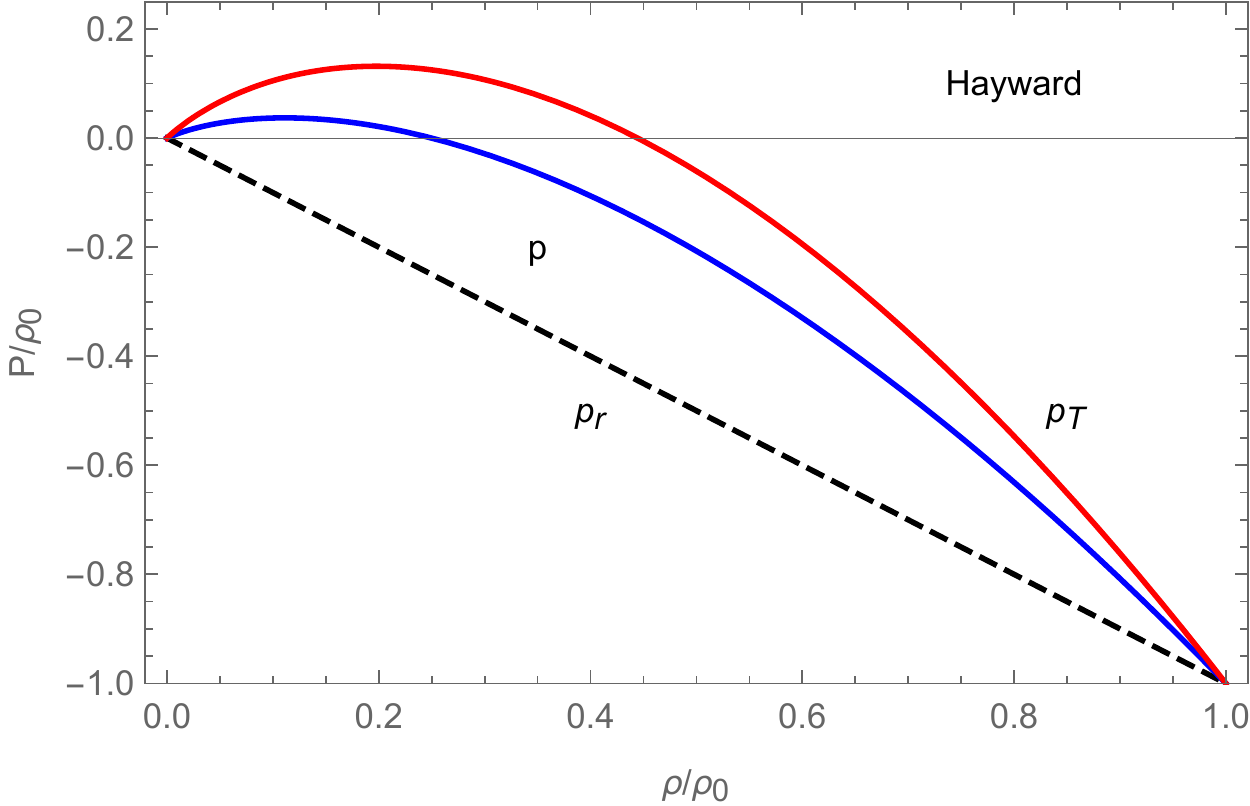}
     \caption{Equation of state plots for the Bardeen, Dymnikova, and Hayward nonsingular black holes. The isotropic pressure $p$ is shown in blue, the transverse pressure $p_T$ is red, and the radial pressure $p_r=-\rho$ is dashed black. Notice all equations of state have a high density point $\rho=\rho_0$ and a low density point $\rho=0$ at which $p=p_T=-\rho$. Also note that in all cases there is a range of lower densities for which the transverse pressure is positive including a smaller range of densities for which the isotropic pressure is positive. }
     \label{bardeeneos}

 \end{figure}

A large number of papers have been published on the variously named solution by Dymnikova originally presented in \cite{Dymnikova1992}. These were derived with Schwinger vacuum polarization for the density \cite{Dymnikova:2000zi}, and also show up under a renormalization scheme of a Schwarzschild black hole \cite{Platania:2019kyx}. The defining functions are
\begin{align}
&   m=\frac{r_g}{2}\left(1-e^\frac{-8\pi \rho_0 r^3}{3 r_g}\right),\\
&   \rho=\rho_0 e^{-\frac{8 \pi \rho_0 r^3}{3 r_g}},\\
&   p_T=-\rho\left(1+\frac{3}{2}\ln\frac{\rho}{\rho_0}\right), \\
&   p=-\rho\left(1+\ln\frac{\rho}{\rho_0}\right),
 \end{align}
where $\rho_0$ and $r_g$ are parameters. Note that the density falls off faster than any power law of $r$, and the dominant energy condition is violated. In fact, the equations of state become infinitely stiff as the density approaches zero. A plot of the equations of state for the Dymnikova spacetime is in Fig. \ref{bardeeneos}.

\end{document}